\documentclass[twocolumn,preprintnumbers,amsmath,amssymb]{revtex4}
\usepackage{graphicx}
\usepackage{dcolumn}
\usepackage{bm}

\begin{document}

\title{Tunneling Recombination in Optically Pumped  Graphene with Electron-Hole Puddles}
\author{
V. Ryzhii$^{1,3}$, M. Ryzhii$^{1,3}$, and T. Otsuji$^{2,3}$} 
\address{
$^1$ Computational Nanoelectronics  Laboratory, University of Aizu,
Aizu-Wakamatsu 965-8580, Japan}
\address
{$^2$ Research Institute for Electrical Communication, Tohoku University, Sendai 980-8577, 
 Japan} 
\address{
$^3$ Japan Science and Technology Agency, CREST, Tokyo 107-0075, Japan
}
\begin{abstract}
We 
evaluate 
 recombination of electrons and holes 
 in optically pumped  
graphene associated with the interband 
tunneling between electron-hole puddles
and calculate the recombination rate and time.
It is demonstrated  that this mechanism
can be dominant in a wide range of pumping intensities.
We show that the tunneling recombination rate and time are  nonmonotonic 
functions of the quasi-Fermi energies of 
electrons and holes and optical pumping intensity. 
This can result in hysteresis phenomena. 
\end{abstract}
\maketitle
\newpage
The gapless energy spectrum of electrons and holes 
in graphene layers (GLs) and non-Bernal stacked multiple graphene layers (MGLs)~\cite{1,2,3}, which leads to  specific features of
its electrical and optical processes, opens up prospects of building of novel devices exploiting the interband
absorption and emission of terahertz (THz) and infrared (IR) photons. Apart from an obvious possibility to use
of graphene in THz and IR photodetectors~\cite{4,5,6}, graphene layers (GLs) and multiple graphene layers (MGLs)
can serve as active media in THz or IR lasers with optical
~\cite{7,8,9,10,11,12} and injection~\cite{13} pumping. 
The achievement of sufficiently strong population inversion, which is necessary for negative dynamic conductivity
in the THz or IR ranges of frequencies, can be complicated by the recombination processes.
The recombination of electrons and holes in GLs  at not too low temperatures is mainly determined by the emission of optical photons~\cite{14}. Due to a relatively high energy of optical phonons ($\hbar\omega_0 \simeq 0.2$~eV), the recombination
rate in GLs 
associated with this mechanism
 can be acceptable even at the room temperatures~\cite{10,11,12}. 
The radiative recombination in the practically interesting temperatures is weaker than that due to the optical phonon emission~\cite{15}.
The acoustic phonon recombination mechanism as well as the Auger mechanism are forbidden due to the linearity
of the electron and hole spectra even in the case their modifications associated with the inter-carrier interaction~\cite{16,17}.
More complex processes involving the electron and hole scattering
on impurities assisted by acoustic phonons also provide rather long
recombination times which are 
longer than  the radiative recombination time~\cite{18}. 
 However in GLs with a disorder caused, for instance, by  fluctuation of the surface charges resulting in the formation of the electron and hole puddles~\cite{19} (see also, for instance, Refs.~\cite{20,21,22}), the recombination can also be
 associated with the interband tunneling in the spots where the build-in fluctuation field is sufficiently strong. 

In this paper, we  find the dependences of
the recombination rate and time in graphene with electron-hole puddles
on the quasi-Fermi energy of electrons and holes
(pumping intensity).
Our results can be used for the interpretation of experimental
observations and to promote the realization of graphene-based THz and IR lasers.

The electric potential $\varphi = \varphi(x,y, z)$, where $x$ and $y$ are the coordinates in the GL plane and the coordinate
$z$ is directed  
perpendicular to this plane, is governed by the Poisson equation presented in the following form:

\begin{equation}\label{eq1}
\Delta \varphi = \frac{4\pi\,e}{\ae} (\Sigma_e - \Sigma_h - \Sigma_i)\cdot \delta(z).
\end{equation}
Here, 
$$
\Sigma_e = \frac{2}{\pi\hbar^2}\int_0^{\infty}\frac{dp\,p}
{1 + \displaystyle\exp\biggl(\frac{v_Wp - \mu_e + e\varphi}{k_BT}\biggr)},
$$
$$
\Sigma_h =\frac{2}{\pi\hbar^2}\int_0^{\infty}\frac{dp\,p}
{1 + \displaystyle\exp\biggl(\frac{v_Wp + \mu_h - e\varphi}{k_BT}\biggr)},
$$
and $\Sigma_i$ are the electron, hole, and charged impurity sheet densities, respectively,
$e = |e|$, $\ae$, $\hbar$, and $k_B$ are the electron charge, dielectric 
constant, reduced Planck's constant,
and Boltzmann's constant,  $v_W \simeq 10^8$~cm/s is the characteristic 
velocity, $p$ is the electron
and hole momentum, $\mu_e$ and $\mu_h$ are the electron and hole quasi-Fermi 
energies counted from the Dirac point, 
$T$ is the temperature, and $\Delta$ is the three dimensional Laplace operator.
The delta function
$\delta(z)$ reflexes the localization of electrons, holes, and impurities in 
the GL plane $z = 0$.
In Eq.~(1), we assumed that the electron spectra or electrons and holes 
are linear: $\varepsilon_e = v_Wp$ and 
$\varepsilon_h = - Wp$. In the equilibrium, i.e., 
in the absence of pumping,
$\mu_e = \mu_h = \mu_i $, where $\mu_i$ is determined by the spatially averaged 
impurity density
$\langle \Sigma_i (x,y)\rangle$ with $\mu_i = 0$ if the latter averaged value 
is equal to zero
(the Fermi levels correspond to the Dirac point) or by the gate voltage
(in gated GL structures). In the case of  pumping of intrinsic GLs, 
$\mu_e = - \mu_h = \mu \neq 0$. 
In the latter case, $T$ can differ from the lattice temperature $T_l$ being both
higher than $T_l$ or lower~\cite{23}.

Considering in the following the case $\langle \Sigma_d (x,y)\rangle = 0$ and assuming that the fluctuations 
 are not too
strong ($e|\varphi| < k_BT$), Eq.~(1) can be linearized:

\begin{equation}\label{eq2}
\Delta \varphi= \biggl[\frac{\varphi}{r_s}\,\ln\biggl(1 +  e^{\displaystyle\mu/k_BT}\biggr)
- \frac{4\pi\,e}{\ae}\Sigma_i\biggr]\cdot \delta(z),
\end{equation}
where
$r_s = (\ae\hbar^2v_W^2/16e^2k_BT)$
is the  screening length.
 
Solving Eq.~(2)
  with the boundary conditions $\varphi|_{z = \pm \infty} = 0$
assuming that
$$
\Sigma_i = \sum_{q_x,q_y}\Sigma^{(i)}_{q_x,q_y}\exp[i(q_xx + q_yy)],
$$
$$
 \varphi(x,y,z) = \sum_{q_x,q_y}\Phi_{q_x.q_y}\exp(-q|z|)\exp[i(q_xx + q_yy)],
 $$
 where $\Sigma^{(i)}_{q_x,q_y}$ and $\Phi_{q_x.q_y}$ are the pertinent amplitudes and
 $q = \sqrt{q_x^2 + q_y^2}$,
 for the amplitude of the potential at the GL plane
 we obtain

\begin{equation}\label{eq3}
\Psi_{q_x,q_y} = \frac{2\pi\,e}
{\ae\biggl[q + \displaystyle\frac{1}{2r_s}
\ln(1 +  e^{\mu/k_BT})\biggr]}\Sigma^{(i)}_{q_x, q_y}.
\end{equation}
Setting $q \sim \pi/{\overline a}$,
where ${\overline a}$ is the characteristic size of the puddles, 
and using Eq.~(3), one
can express $\Psi = max |\Psi_{q_x,q_y}|$ via its value
in the absence of screening ${\overline \Psi} = 
2e{\overline a}{\overline \Sigma}/\ae$:

\begin{equation}\label{eq4}
\Psi = \frac{{\overline \Psi}}
{\biggl[1 + \displaystyle\frac{{\overline a}}{2\pi\,r_s}\,
\ln(1 +  e^{\mu/k_BT})\biggr]}.
\end{equation}

According to Eq.~(4), the fluctuation electric field  between positively
and negatively charged puddles can be estimated as
${\cal E} \sim \Psi/{\overline a}$.

\begin{figure}[t]
\center{\includegraphics[width=6.5cm]{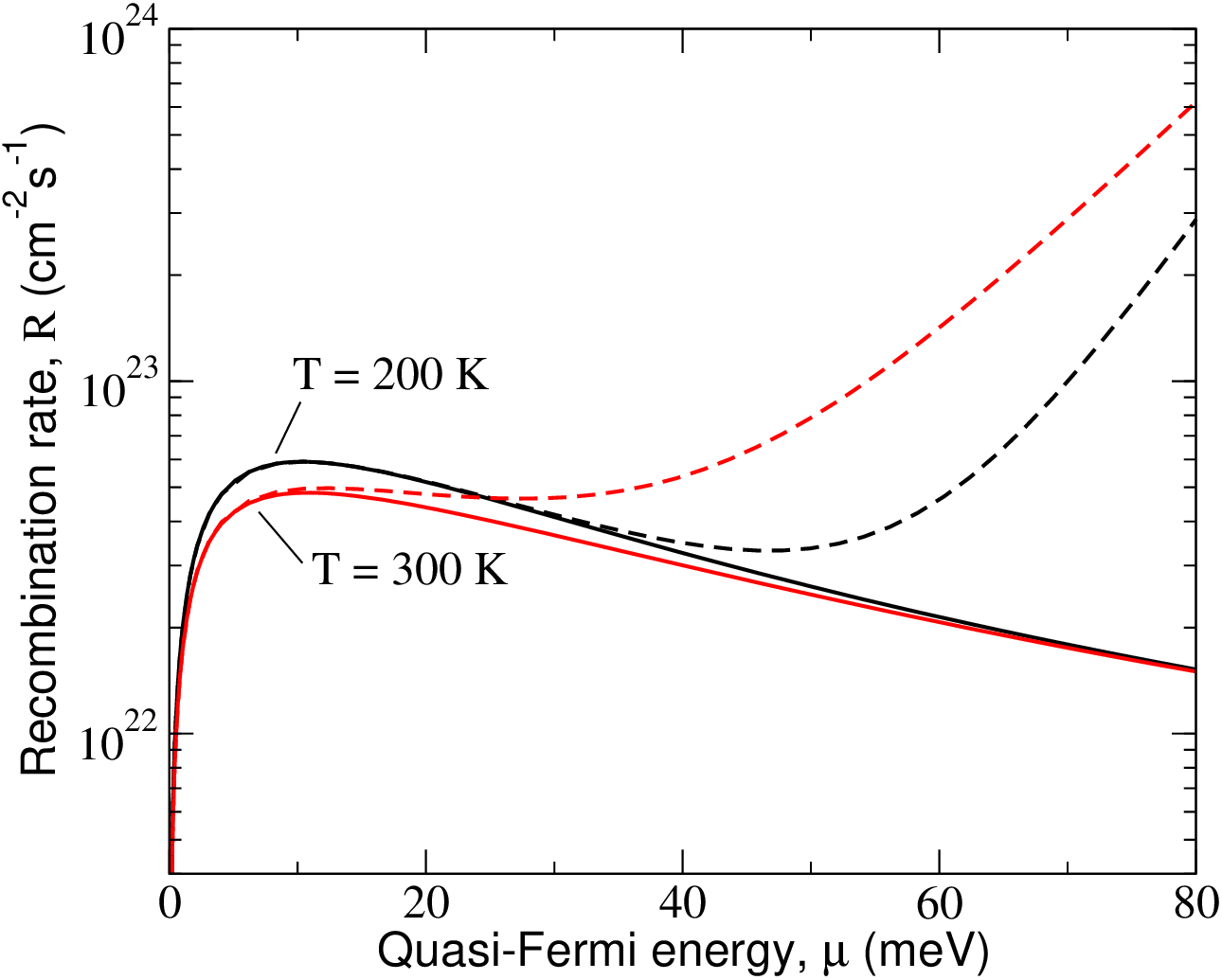}}
\caption{
(Color online) Recombination rate $R$ vs quasi-Fermi energy $\mu$ 
for tunneling mechanism (solid lines) and for combination of tunneling and optical phonon mechanism (dashed lines)
in a GL  at  different temperatures $T$. 
}
\label{Fig1}
\end{figure}

\begin{figure}[t]
\center{\includegraphics[width=6.5cm]{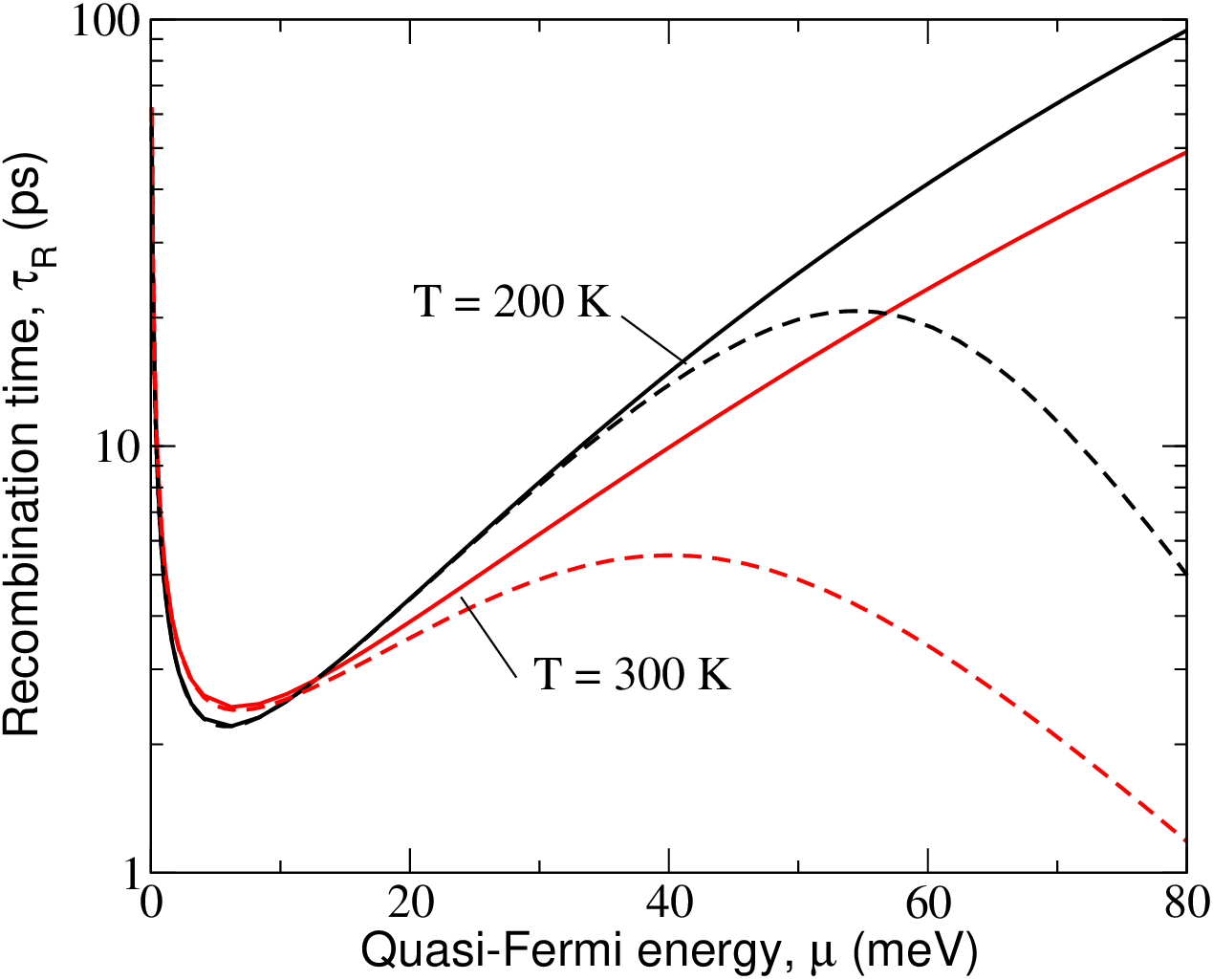}}
\caption{
(Color online) Recombination time $\tau_R$  vs quasi-Fermi energy $\mu$ at different temperatures $T$. Solid lines correspond to tunneling recombination,
while  dashed lines correspond
to recombination associated with both tunneling and phonon mechanisms.
}
\label{Fig2}
\end{figure}

\begin{figure}[t]
\center{\includegraphics[width=6.5cm]{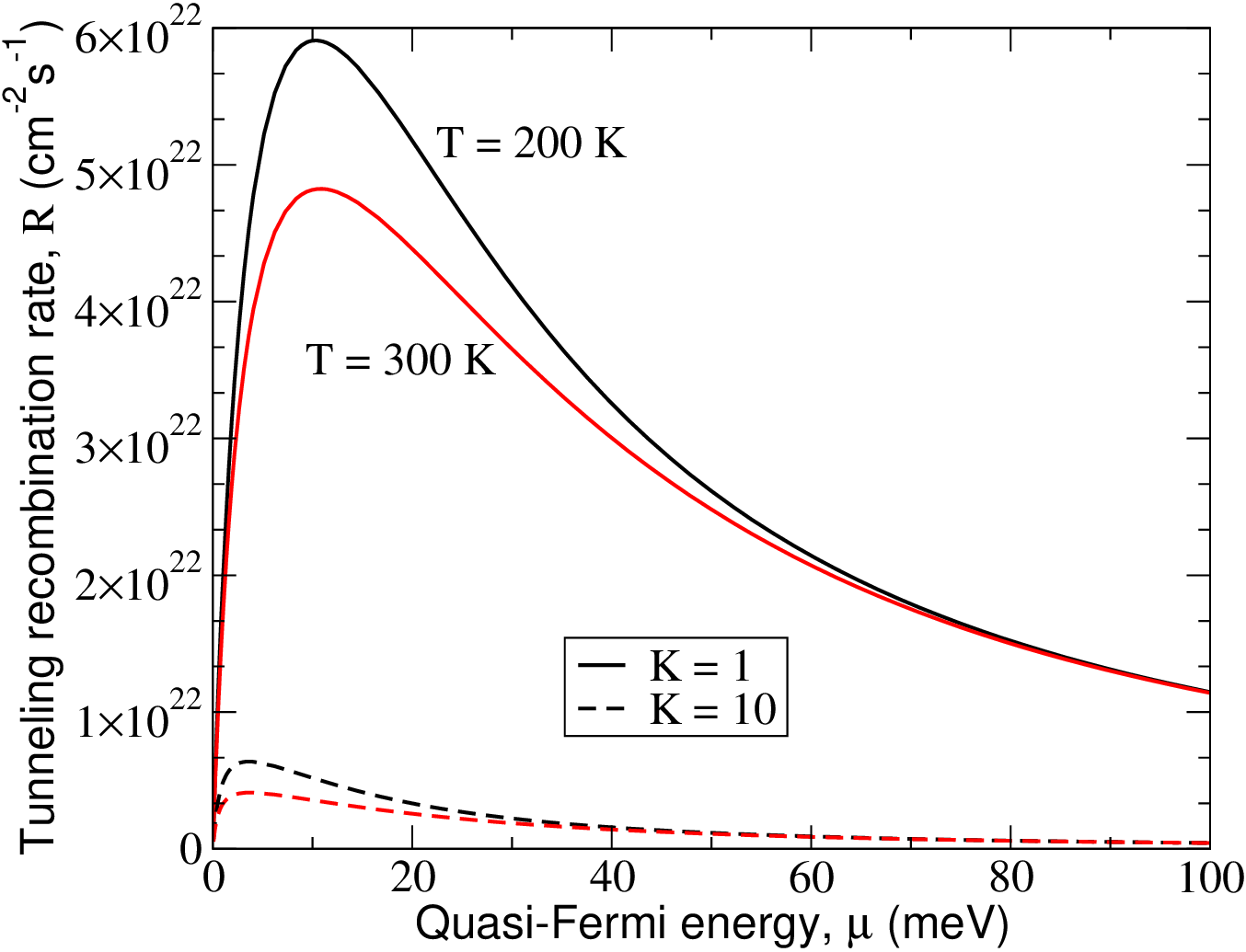}}
\caption{
(Color online) Tunneling recombination
 time $\tau_R$  vs quasi-Fermi energy $\mu$ at different temperatures $T$
in a GL (solid lines) and in MGL structure with ten GLs, i.e.,$K = 10$ (dashed lines).
}
\label{Fig3}
\end{figure}

Using the general formulas for the probability of
the interband tunneling in GLs~\cite{24,25,26} and considering Eq.~(4),
the rate of the tunneling recombination
 can be presented  as
\begin{equation}\label{eq5}
R \sim \frac{2\sqrt{2}e^{3/2}}{\pi^2\hbar^{3/2}v_W^{1/2}}
\sqrt{\frac{\Psi}{{\overline a}}}
\biggl(\frac{\mu}{e{\overline a}}\biggr) =  \frac{{\overline R}\,(\mu/e{\overline \Psi})}
{\sqrt{1 + \displaystyle\eta_s
\ln(1 +  e^{\mu/k_BT})}},
\end{equation}
\begin{equation}\label{eq6}
R \sim \frac{2\sqrt{2}e^{3/2}}{\pi^2\hbar^{3/2}v_W^{1/2}}
\biggl(\frac{\Psi}{{\overline a}}\biggr)^{3/2}=  
\frac{{\overline R}}
{[1 + \displaystyle\eta_s
\ln(1 +  e^{\mu/k_BT})]^{3/2}}
\end{equation}
 at relatively weak  ($\mu < e\Psi$) and strong ($\mu >  e\Psi$) 
 optical pumping, respectively.
Here,
$$
{\overline R} \sim \frac{2\sqrt{2}e^{3/2}}{\pi^2\hbar^{3/2}v_W^{1/2}}
\biggl(\frac{{\overline \Psi}}{{\overline a}}\biggr)^{3/2}, \qquad \eta_s 
=\frac{{\overline a}}{2\pi\,r_s} = 
\frac{16}{\pi\ae}\frac{e^2{\overline a}k_BT}{\hbar^2v_W^2}.
$$
Assuming that $\ae = 4$, ${\overline a} = 30$~nm, and ${\overline \Sigma} = 
4\times10^{10}$~cm$^{-2}$~\cite{19}, for the amplitude of the potential profile fluctuations
$e{\overline \Psi}$ and for the characteristic rate of the tunneling recombination
${\overline R}$, we obtain the following estimates:  $e{\overline \Psi} = 8.5$~meV
and
${\overline R} \simeq~ 2.7\times 10^{23}$~cm$^{-2}$s$^{-1}$, respectively.
At  $T = 77 - 300$~K, one obtains  $r_s \simeq 0.67 - 2.62$~nm. These values are used in the following.

In the case of  strong pumping, the effective temperature of the electron-hole plasma 
can be much higher than the lattice temperature~\cite{23}.
As a result,
 the electron-hole plasma can become nondegenerate,
so that $\mu_e = \mu < 0$ while $\mu_h = - \mu  > 0$. This is possible when the plasma density increases with increasing pumping intensity 
slower that the effective temperature rises.
In such a case, the screening vanishes ($\Psi \simeq {\overline \Psi}$), and $R < 0$. This implies that when $\mu$ changes its sign, the tunneling
recombination of  electron-hole pairs turns to their tunneling generation.
This is because when the plasma density is lower than it would be in 
equilibrium but at the effective temperature $T$, the tunneling generation
tends to establish an equilibrium density corresponding to $T$.

The dependences of the tunneling recombination rate $R$ versus the quasi-Fermi energy $\mu$
for different temperatures $T$ are  shown in Fig.~1.
For the calculations we used a formula for $R$ which  interpolates the dependences given by Eqs.~(5)
and (6). Figure~1 demonstrates also the dependences of the recombination rate
$R + R_0$
associated with both tunneling and optical phonon mechanisms (dashed lines).
The contribution of the optical phonon mechanism $R_0$ is taken into account
using a simplified  analytical formula~\cite{10,23} 
derived from more rigorous one~\cite{14}:

$$
R_{0}
= {\overline R_0}\biggl[
{(\cal N}_0 + 1)\exp\biggl(\frac{2\mu - \hbar\omega_0}{T}\biggr)  
- {\cal N}_0
\biggr] 
$$
\begin{equation}\label{7}
\simeq {\overline R_0}\exp\biggl(-\frac{\hbar\omega_0}{k_BT}\biggr)
\biggl[\exp\biggl(-\frac{2\mu}{k_BT}\biggr) - 1\biggr]. 
\end{equation}
Here ${\overline R_0} \simeq 10^{23}$~cm$^{-2}$s$^{-1}$ ~\cite{14} 
is the pertinent 
characteristic recombination rate 
and ${\cal N}_0 = [1 +\exp(\hbar\omega_0/k_BT)]$ 
is the number of optical phonons.
Disregarding the effects of  electron-hole heating and cooling and the effect 
of optical phonon 
heating~\cite{23}, we have neglected the difference in the effective temperature $T$ and the lattice 
temperature $T_l$.
One can see that $R$ is a nonmonotonic function of $\mu$  with a maximum 
at a certain value of $\mu$. This is attributed to an increase in $R$ at small $\mu$
due to an increase in   the electro and hole densities, i.e. in the number of carriers participating in the tunneling processes. However, 
at higher values of $\mu$, the screening of the potential fluctuations leads
to a decrease in the electric field between the puddles and, consequently,
in a decrease in the tunneling probability. When $\mu$ increases further,
the tunneling mechanism gives way to 
the optical phonon mechanism which  becomes dominating. 
This results in a dramatic
rise of $R + R_0$ in the range of large $\mu$.
The quantity $\mu$ is governed by an  equation equalizing
the rate of recombination  $R + R_0$ and the rate of carrier photogeneration
$G = \beta I$, where $\beta = 0.023$ is the GL absorption coefficient
and $I$ is the photon flux. Hence, the $\mu - I$ dependence
can   exhibit the S-shape (see the dashed line in Fig.~1
 corresponding to $T = 200$~K) leading to a hysteresis
which might be pronounced
at lowered temperatures.  

The recombination time is given by  
$$
\tau_R = \frac{\langle(\Sigma_e + \Sigma_h)\rangle}{2(R + R_0)}
$$
\begin{equation}\label{8}
= \frac{2}{\pi\,(R + R_0)}\biggl(\frac{k_BT}{\hbar\,v_W}\biggr)^2\int_0^{\infty}\frac{d\varepsilon\,\varepsilon}
{1 + \displaystyle\exp(\varepsilon  - \mu/k_BT)}.
\end{equation}
The recombination time calculated without considering
optical phonon 
mechanisms  (solid lines) and with the latter (dashed lines) 
is shown in Fig.~2.
As follows from Fig.~2, the $\tau_R - \mu$ dependence is also nonmonotonic 
with a minimum at moderate $\mu$ (where $\tau_R \sim 2$~ps) and a muximum
at rather high $\mu$ (where $\tau_R \sim 5 - 20$~ps).

In optically pumped MGL structures, the electron and hole densities in each GL
are close to each other provided the number of GLs $K$ is not too 
large~\cite{27}.
If the fluctuations are associated with the charges located near the
lowermost GL, say, near the interface between this GL and the substrate,
these fluctuations are screened by all GLs.
In cases of  such MGL structures, the first two terms in the right-hand
side of Eq.~(1) can be multiplied by a factor $K$, so that the screening
length becomes much shorter:
$r_s^{(K)} = (\ae\hbar^2v_W^2/16Ke^2k_BT)$. Consequently, parameter
$\eta_s$ should be substituted by $K\eta_s$
Naturally, this leads to a stronger suppression of the fluctuations
(by a stronger screening) and, hence to weakening of the 
tunneling recombination under consideration as shown in Fig.~3.
If each GLs contains its own fluctuation charges,
the averaging over a number of GL leads to a significant smoothening
of the potential relief in all GL and, hence, to the suppression of the 
tunneling recombination is such MGL structures. 

In conclusion, we calculated the recombination and 
time as functions of the quasi-Fermi energy in optically pumped
graphene with electron-hole puddles. It was shown that the tunneling recombination can be dominant recombination mechanism at low and moderate values of 
quasi-Fermi energy and, hence, at low and moderate pumping intensities.
The dependences of the recombination and time on the quasi-Fermi energy and
pumping intensity can be nonmonotonic resulting in hysteresis phenomena.
The tunneling recombination can be diminished  in MGL structures.

The authors are grateful to N.~Ryabova for comments on the manuscript
and S. Boubanga Tombet for useful information.
This work was supported by the Japan Science and Technology Agency, CREST and by the Japan Society
for Promotion of Science, Japan.

\end{document}